\documentclass{tcibook}
\usepackage{fancyhea}
\usepackage{work}
\usepackage{bm}       
\usepackage{graphicx}
\usepackage{multirow}
\usepackage{lineno}
\usepackage{threeparttable}
\usepackage[normalem]{ulem}
\usepackage{color}
\usepackage[usenames]{xcolor} 

\usepackage[colorlinks = true,
            linkcolor = blue,
            urlcolor  = blue,
            citecolor = blue,
            anchorcolor = blue]{hyperref}
            
\def\beq{\begin{equation}}
\def\eeq#1{\label{#1}\end{equation}}
\def\eeqn{\end{equation}}

\newenvironment{Eqnarray}%
   {\arraycolsep 0.14em\begin{eqnarray}}{\end{eqnarray}}
\def\beqa{\begin{Eqnarray}}
\def\eeqa#1{\label{#1}\end{Eqnarray}}
\def\eeqan{\end{Eqnarray}}

\let\bar=\overbar

\def\lsim{\mathrel{\raise.3ex\hbox{$<$\kern-.75em\lower1ex\hbox{$\sim$}}}}
\def\gsim{\mathrel{\raise.3ex\hbox{$>$\kern-.75em\lower1ex\hbox{$\sim$}}}}

\def\del{\partial}
\def\Dslash{\not{\hbox{\kern-4pt $D$}}}
\def\dslash{\not{\hbox{\kern-2pt $\del$}}}
\def\pslash{\not{\hbox{\kern-2pt $p$}}}
\def\ETmiss{\not{\hbox{\kern-4pt $E$}}_T}

\def\Dlr{\mathrel{\raise1.5ex\hbox{$\leftrightarrow$\kern-1em\lower1.5ex\hbox{$D$}}}}

\def\MSB{{\bar{M \kern -2pt S}}}
\def\msb{{\bar{\scriptsize M \kern -1pt S}}}

\def\drb{{\bar{\scriptsize D \kern -1pt R}}}

\def\authorlist#1#2{
    \vskip 0.4in
\begin{center}\begin{large} {\bf  #1 } \end{large}
    \vskip 0.2in
              #2
     \vskip 0.2in
   \end{center}
}

\begin{document}


\pagenumbering{roman}

\parindent=0pt
\parskip=8pt
\setlength{\evensidemargin}{0pt}
\setlength{\oddsidemargin}{0pt}
\setlength{\marginparsep}{0.0in}
\setlength{\marginparwidth}{0.0in}
\marginparpush=0pt


\pagenumbering{arabic}

\renewcommand{\chapname}{chap:intro_}
\renewcommand{\chapterdir}{.}
\renewcommand{\arraystretch}{1.25}
\addtolength{\arraycolsep}{-3pt}

\newcommand{\EVHcomments}[1]{\textcolor{cyan}{#1}}
\newcommand{\EScomments}[1]{\textbf{\textcolor{purple}{#1}}}
\newcommand{\comments}[1]{\textbf{\textcolor{violet}{#1}}}

\newcommand{\cefgroup}{3} 
\newenvironment{recs}
    {\noindent \rule{\textwidth}{2mm} }
    {\rule{\textwidth}{2mm}}

\newcommand{\rec}[3]{
\begin{minipage}{\textwidth}
\vspace{0.05in}
\nobreak \noindent \textbf{CEF0\cefgroup~Recommendation #1 -- #2}\\ 
\nobreak \rule{\textwidth}{0.4mm}
\end{minipage}
\nobreak #3\\
}

\setcounter{chapter}{2} 


\chapter{Diversity, Equity, and Inclusion\\ in Particle Physics}

\authorlist{C. Bonifazi,  J. S. Bonilla, M.-C. Chen, Y. H. Lin}
   {K. A. Assamagan, E. V. Hansen, S. Meehan, E. Smith}

\section{Executive Summary}   
\label{sec:executive_summary}

To achieve the highest level of intellectual excellence calls for the greatest extent of diversity. However, due to the unjust institutional and societal barriers, the field of particle physics remains as one of the least diverse fields, severely limiting the potential of our scientific achievements. In order for the US Particle Physics Community, including the accelerator science and engineering fields, to remain at the forefront of global scientific leadership, it is imperative for our community to act urgently and diligently to improve the status quo of diversity, equity, inclusion, and accessibility (DEIA).

In order to improve the DEIA in particle physics to further our scientific goals, we must allocate dedicated financial and personnel resources to 
\begin{itemize}
    \item Bring awareness in the particle physics community about different forms of marginalization, including but not limited to racism experienced by individuals identified as Black, Hispanic or Latino/a/x, Indigenous, Asian, as well as other forms of discrimination based on gender identities, disability status (both visible and invisible) and neurodiversity; LGBTQA+; veteran status; socio-economic status, xenophobia, and intersectionality of these identities. Educate our community to be good bystanders.
     
     \item \textbf{Create pathways} for members from historically and currently marginalized backgrounds to participate in particle physics community, and provide necessary support (including but not limited to accessibility, personal, financial) for these members to have equitable opportunity to thrive in our field. 
     
     \item \textbf{Engage communities} from emerging and developing countries, including Africa and Latin America, to ensure DEIA in the participation of members from these regions in our global endeavors of particle physics. 
     
     \item \textbf{Engage outside experts} (including sociologists and psychologists) to help develop effective strategies for continuous improvements, through 
     
     \begin{itemize}
     \item effective engagement with marginalized communities to increase their representation;  \item fostering an inclusive climate in our field, so that all members feel welcome and supported, both socially and financially in their academic and  research participation;
     \item establishing equitable evaluation, selection, and reward processes by interrupting implicit bias and recognizing ``unearned privileges'' to ensure every one of our community members has equal opportunity for advancement.
     \end{itemize}
     
     \item \textbf{Restructure the culture} of particle physics to attain work-life balance. Bring awareness about and support members with mental health conditions.
  
     \item \textbf{Enact the agency} in each individual community member to work towards DEIA. Institute policy changes to hold individual and institutions accountable. Create structure to systemically monitor the progress of DEIA in our field.
     
\end{itemize}

All members in the particle physics community would benefit from such cultural and policy changes, both through having a healthy research and education environment that is conducive to fostering long-term, sustained productivity, and through the diverse talents said environment will attract to maximally enhance our scientific excellence.


\section{Introduction}
\label{sec:intro}

Diversity, equity, inclusion, and accessibility (DEIA) is necessary for maximizing the intellectual excellence in particle physics. Due to institutional barriers and societal biases, not all members in our community have equal access to education or career and advancement opportunities. This results in the lack of diversity in our field, and thus limits the potential of advancement that can be achieved. We do not have a choice of what circumstances we were born into, but we do have a choice to shape the future.

Our world has precedent and prejudice; their effect permeates through every professional sphere, physics is no exception. It is our responsibility as academics to compensate for these pressures in order to provide each and all of us with the space needed to fully contribute to the advancement of our field. The reality, however, is that we often fall short of providing such compensation. It is important for our community to recognize that until our community representation is diversified, and members from all identities are afforded an environment where the climate and support systems are in place to ensure their success, our scientific achievements will continue to be limited.

\section{CP Summaries}
\label{sec:cpsummary}

Diversity, Equity, and Inclusion encompassed a broad range of aspects. Over the two-year long process of Snowmass 2021, the Community Engagement Frontier Diversity, Equity, and Inclusivity (DE\&I) Topical Group (CEF3) has hosted 12 community town halls both to engage the community in the discussions of DEI issues and to support our community members  during the pandemic. Letters of Interest were solicited and 29 were received with CEF3 tags. Snowmass 2021 has resulted in 11 white papers on a variety of topics. We summarize the findings of these white papers in this Section.


\subsection{Experiences of Marginalized Communities in HEP~\cite{HowToReadSelectedSnowmassPapers, PowerDynamicsInPhysics, PolicingAndGatekeepingInSTEM, InformalSocializationInPhysicsTraining,
MarginalizedCommunities}}   
\label{section:experiences}


Social structures and interactions in physics in power dynamics, informal socialization, policing and gate keeping, affect negatively the experiences and participation of underrepresented people, with a lack of awareness and attention---or a focus on perception rather than reality---from those with privilege. Barriers and exclusionary practices influence the experiences of people with marginalized identities and challenge the claims of objectivity in physics. Such claims only serve to re-enforce the lack of awareness and focus on perception by the privileged community and exacerbate marginalization of people with underrepresented identity. Perception of objectivity, supported by barriers and exclusionary systems, deny the experiences of physicists with marginalized identified, with the expectation---from privilege community---that marginalized physicists should shrug off their negative experiences, despite the harms caused, to be taken seriously~\cite{HowToReadSelectedSnowmassPapers, PowerDynamicsInPhysics, PolicingAndGatekeepingInSTEM, InformalSocializationInPhysicsTraining}. The key to improve the experiences of BIPOC physicists consists of addressing the complexity and impact of power dynamics, policing and gate keeping~\cite{PowerDynamicsInPhysics, PolicingAndGatekeepingInSTEM}, and implementing actions individuals and organizations can take to lower barriers for early career BIPOC physicists~\cite{InformalSocializationInPhysicsTraining}.

Although the particle physics community has devised programs to improve diversity and inclusion along multiple axes, and the way that we think about measuring and implementing inclusion initiatives has evolved, the aforementioned issues remain as evidenced in the personal experiences and testimonials described in Refs.~\cite{HowToReadSelectedSnowmassPapers, PowerDynamicsInPhysics, PolicingAndGatekeepingInSTEM, InformalSocializationInPhysicsTraining, MarginalizedCommunities}. In Ref.~\cite{MarginalizedCommunities}, the authors outline steps to build a culture of equitable access for the success of marginalized members. Specifically, urgent needs are identified in the following areas: (1) acquire a better understanding of the status quo, both quantitatively and qualitatively, in order to develop best practices to improve the experiences of marginalized members and to assess the effectiveness of the existing programs; (2) develop effective and inclusive ways to engage marginalized communities and provide training to community members to implement the engagements; (3) create infrastructure to better support members of marginalized communities scholarly, financially, and personally; (4) create an environment for access and success of marginalized members by establishing community expectations, fostering inclusion in social interactions, and holding individuals and institutions accountable for their actions; (5) establish a mechanism to monitor the progress towards DEIA, including the implementation of recommendations resulting from the Snowmass 2021 process.  


\subsection{In Search of Excellence and Equity in Physics~\cite{ExcellenceAndEquityInPhysics}}   
\label{section:equityandexcellence}
To increase equal opportunities in physics, we must ensure that the leaders chosen for institutions are based on merit and not connections.  A fair evaluation of the candidates for the high-level leadership can mean healthier workplace climate and fair evaluation of candidates for all other positions within the organization.  Institutions must show that discrimination in the workplace will not be tolerated and that changes are not superficial.  Many things can ensure a more diverse candidate pool and more impartial hiring process: performing a job analysis, advertising the leadership positions broadly (particularly to underrepresented professional societies), assembling a diverse search committee who are trained on recognizing and mitigating their own biases, establishing rubric before the process begins, using a structured and consistent process across applicants, and asking the candidates to react to a EDI-related problem case study.  The process should be transparent and equitable.  This will ensure more trust from the stakeholders, funding agencies, and the organization in the selected leader.

\subsection{Accessibility in High Energy Physics: Lessons from the Snowmass Process~\cite{AccessibilityInHEP}}   
\label{section:accessibility}

The physics community must actively protect people’s fundamental right to participate in physics regardless of disability, identity, or background.  Barriers, defined as anything that prevents or makes substantially more difficult to actively participate in physics activities.  Oftentimes, an individual can experience multiple barriers, which can be more than just the sum of the parts.  An accessibility survey of the Snowmass community has shown the majority of respondents reporting experiencing financial, caretaker responsibilities, mental health, physical, auditory, and visual barriers of varying degrees for either themselves and/or other around them.  

There is a dichotomy of perceived and real challenges of the these barriers, particularly auditory barriers, as more meetings become virtual.  While transcripts and auto-captioning services are available, they can lack the ability to recognize specialized jargon, have training bias towards English-speaking white males, and not pick up on sarcasm or emotions, which all contribute to inability to participate in a conversation.  

While most of the respondents rely on home institutes or their personal/group grants to provide access currently, most agree that American Physical Society should be responsible for providing support and that the burden should not be the individual.  There are certainly resources available, though they often require advance planning and application for the funds.  

The key logistical recommendations are planning ahead and budgeting appropriately for accommodations, having a single point of contact for ease of coordination, educating yourself about etiquette by consulting resources, along with clear and transparent communication throughout the process.  One should not assume that because one accommodation works for an individual that it will work for all individuals with similar barriers; after all, barriers come in varying levels and the intersectionality between them can introduce other complications.  The types of accommodations that work best for each individual depend on their context and experiences.

\subsection{Climate of the Field: Snowmass 2021~\cite{ClimateOfTheField}}   
\label{section:climate}


Racism, misogyny, and other fundamentally oppressive, violent, and exclusionary ideas and practices exist in scientific research environments as they do in society. When developing community-wide policies it is vital to seek out best practices developed by experts, with careful consideration of who we regard as experts~\cite{ExcellenceAndEquityInPhysics}, to ensure these policies are effective and not harmful. To this end, this contributed paper provides recommendations for several top-down approaches that should be implemented by the community as well as recommendations for funding agencies to support these approaches. First, physicists should not be doing the job of organizational psychologists and DEI experts; instead, funding should be set aside to engage with and compensate experts. Second, our communities should be asked regularly how their leadership is designed --- how their affiliates are trained and selected for leadership roles. Finally, funding and structural aid should be made available to develop formal structures at host laboratories to provide the HEPA community with advice and logistical tools related to DEI and Code of Conduct issues. 

Interest groups, professional organizations, and scientific collaborations with member institutions from all over the world have regular in-person meetings at single participating institutions. Generally, these groups have developed a code of conduct that sets expectations for behavior at these meetings, and often, potential sanctions for members who violate the code of conduct. However, many of these groups --- particularly scientific collaborations --- operate in a gray area as they are not a legal entity and often do not have oversight or the resources to create that oversight. As a result, many organizations' codes of conducts are rendered useless. To illustrate, consider an incident an in-person meeting of a collaboration. The likelihood that this involves community members who are not members of the meeting's host institution is high. Under existing policies, the host institution is unlikely to investigate if neither collaboration affiliate is associated with that institution, and the victim’s and the perpetrator’s institutions are unlikely to investigate incidents that do not occur on their campus. One might look to the experiment's host laboratory to investigate as it is the common factor between collaboration affiliates; however, many labs will not investigate if neither affiliate is an employee, and while some labs state that they treat issues between visitors in the same way as issues between employees, this is not consistent across all labs and investigations of harassment should not depend on the generosity of a particular institution. 
These investigations may be left to HR or counsel, but experiences with these offices leave the impression that their objective is to protect the institution, and that resolutions to such cases prioritize the institution over victims. 
Having exhausted these possibilities, one might look to collaboration leadership to enforce the code of conduct; however, physicists are not trained nor equipped to perform investigations misconduct, and collaborations do not have funding allocated for an independent third party investigator, nor are there investigators on retainer at institutions. Because collaborations are not legal entities themselves, any sanctions taken against someone engaging in misconduct leaves collaboration affiliates who enact these sanctions open to legal action brought against them by the perpetrator. The inability to enforce our codes of conduct represents a huge failure of our institutions to provide a safe working environment. To change this, we must develop community-wide policies that go beyond symbolic legal compliance, which protect institutions from legal liability but do not prevent harassment. Our institutions must develop ``clear, accessible, and consistent policies'' on standards of behavior which should include ``a range of clearly stated, appropriate, and escalating disciplinary consequences'' for those found to have violated policies and/or law, 
and these policies should be consistent across subfields and experiments. We cannot develop codes of conduct without being prepared to exercise them. If we are not prepared to exercise them, it is better not to have them at all. 

This contributed paper includes interviews of several people that have an equity-, diversity-, and inclusion-related role in their collaboration. From these interviews several common themes and pitfalls emerged regarding codes of conduct, including common pushbacks from collaborators before and during development of a code of conduct and before adoption, difficulties in developing a code of conduct without institutional support, difficulties in identifying appropriate mechanisms for reporting and investigating violations. Additionally, those in EDI roles were aware that there are best practices for developing codes of conduct but were unsure how to find them; a great deal of time and energy was spent ``reinventing the wheel.'' 

This contributed paper also addresses the inequities across identity groups, using the roles of scientists and nonscientists (e.g. accelerator technicians) as an example. These groups are not equally included in collaborations, from inequities in authorship to exclusion from communication channels or social events. Young technicians often receive less support, inclusion, mentorship, and recognition. Recommendations for imitation game activities and mirroring this scientist-nonscientist relationship into other minoritized identity groups is also presented. 

Finally, this contributed paper addresses inequities in information sharing and onboarding, specifically regarding software. Interviews with former community members are presented, with particular focus on frustrations with communication and information transfer between junior and senior scientists. For example, documentation is often the last thing to be written, and relies heavily on graduate students and postdoctoral researchers to develop --- junior scientists who are not trained as teachers or mentors. This results in incoming collaborators spending \textit{years} just to catch up with an evolving and changing framework. ``This is just how things are" has never been acceptable, and changes must be made to code development, documentation, and information transfer. 

\subsection{Lifestyle and personal wellness in particle physics research activities~\cite{LifestyleAndPersonalWellness}}   
\label{section:lifestyle}
The HEPAC community is immersed in a very competitive environment, such that often individuals end up working after hours and during the weekends, affecting the work-life balance, with consequences on mental health. This situation is enhanced for students, postdocs and young researchers who have to compete for a position. Working from home is more impacted by the living conditions, therefore working off office hours enhances the inequity. A key problem that affects our field (and the society in general) is gender imbalance. Among a long list of examples of gender imbalance observed in our field, we can mention the existing differences in salaries for the same position or kind of work, the imbalance in key-power positions, or the role in the care of others. 

Another issue is that some activities that are key for scientific life are either not counted as part of the job or are non-payable duties. Some examples are outreach, EDI initiatives or mentoring that are mainly not counted as part of the job and individuals that spent part of their time on those activities are not compensated for that. Journal refereeing, project reviews, and hiring committees are considered additional non-payable duties. This puts pressure on people to do this type of activity outside of working hours. Furthermore, in many cases, these types of activities end up overburdening the most underrepresented groups. Researchers in the early stages of their careers are often paid low salaries and have few resources to carry out their tasks. The problem is that these people are expected to be high performers and often the personal and family difficulties they may have at this point in their careers are not taken into account.

Despite all the actions and programs to diminish discrimination and harassment at the universities and national labs, such as codes of conduct and ombudspersons, individuals are still being affected by this kind of aggression. A more subtle and, somehow, less evident way of degrading and insulting individuals is what we know by microaggressions. This kind of stereotyping, exclusion and/or sarcastic comments toward stigmatized or culturally marginalized groups produce a very hostile work environment. 

The problems pointed out above cause numerous effects on mental and physical health that affect individuals at all levels. This became a recursive process that feeds back causing an increase of the gaps. A new framework of collaborative work is urgently needed for the community and for that the most complete overview of the situation should be analyzed. A set of actions should be taken at the different levels to improve the situation and to build a sustainable, healthy and equitable work environment for the members of our community. 


\subsection{Why should the U.S. care about high energy physics in Africa and Latin America?~\cite{HEPInAfricaAndLatinAmerica}}   
\label{section:emerging}
Research and education activities in high energy physics are worldwide efforts; however, the contributions from developing countries in Africa and Latin America are severely restricted by available resources and national priorities. Aids for international developments serve the U.S. national agenda and global competitive edge, as defined through the Title VI Act---this is the main reason for the U.S. to maintain or increase support for physics research and education in Africa and Latin America. High energy physics, through its international activities, can help serve these U.S. interests while fostering sustainable development for improved quality of life for all people~\cite{HEPInAfricaAndLatinAmerica}.

There are many efforts over the years to provide scholarships, travel grants, fellowships, etc. to scientists from countries in development, but the major obstacles keeping these countries from fully participating in our global quest for physics have not been dismantled. Many scientists from countries in development do not have a home department with funding, open positions, or a supporting culture to as easily create a HEP group when one doesn't exist. Furthermore, entrance fees and commitments required to join an existing scientific collaboration are often agnostic to the capacity or development stage of the applying institute's host country. More specifically, if an established R1 American institute wishes to join an LHC experiment, the membership fees and maintenance dues are the same as that of a burgeoning group in Latin-America or Africa whose available funding is orders of magnitude less than that of the US. This CP elaborates on this and other systemic obstacles that are preventing the full participation of all scientists around the globe. Recommendations will be provided in Section \ref{sec:recommendations}.
\subsection{Strategies in Education, Outreach, and Inclusion to Enhance the US Workforce in Accelerator Science and Engineering~\cite{StrategiesToEnhanceAcceleratorWorkforce}}   

\label{section:accelerators}

Despite the growing number of women in attendance of the US Particle Accelerator School (USPAS), this growth is not reflected in the percentage of woman at Department of Energy national labs.  Many well-understood barriers contribute to this hindrance which negatively impact the inclusion of gender diversity in the accelerator field.

Similarly, the counts of Black/African American or Native American ancestry are very low, based on the demographic data from the last three years.  Prior to 2021, USPAS did not collect race or ethnicity data of students consistent with URM classification. 

Many national labs and institutions have taken the initiative of creating programs that support the hiring and retention of marginalized populations of gender and historically URM.  However, more work needs to be done to understand the intersection between these underrepresented identities.  Having a consistent demographic data collection of the field will help to identify shortcomings in the recruitment process and best practices in the retention efforts.  

 Overall, this paper recommends the following for Accelerator Science and Engineering:
 \begin{itemize}
     \item National labs and institutions can follow the lead of the American Physics Society Inclusion, Diversity, and Equity Alliance (IDEA) initiative and create programs that will recruit and promotion the retention of marginalized populations in the field
     \item Casting a wide net when hiring through posting on public channels, rather than relying on the network of the hiring manager.  Build network with professional societies promoting the well-beings of marginalized populations
     \item Consistently collect demographic data to monitor the outcome of the recruitment and retention efforts
     \item Reward recruiting and outreach efforts, diverse hiring, and building an inclusive environment to encourage retention
 \end{itemize}
\section{Outlook}
\label{sec:outlook}
Progress towards diversity, equity, inclusion and accessibility requires sustained and committed efforts. During the current Snowmass 2021 process, the current status quo on DEIA has been evaluated, issues have been raised, and some strategies have been recommended. This is an initial step that should be followed up and sustained. 

By the next Snowmass process, we hope to see improvement to the overall climate of particle physics. We hope to see particle physics become more diverse, equitable, inclusive and accessible. The progress should be reflected both in the diversity of representation at all career stages, and in the experiences by all members in our community.  We also hope to see statistics on how many of the recommendations outlined in this and other white papers have been implemented. 

Finally, as has been pointed out in this and other Snowmass white papers, and has been demonstrated in the process of writing those white papers, DEIA-focused tasks often disproportionately fall upon people from marginalized groups. In the next Snowmass process, we hope that the demographics of participants in DEIA-related activities will better reflect the demographic distribution of all Snowmass participants. In addition, we hope that our community will soon establish an entity, possibly within the Division of Particles and Fields, to monitor the progress toward diversity, equity, inclusion, and accessibility.

\clearpage
\section{Recommendations}
\label{sec:recommendations}
We summarize here the recommendations compiled as a result of the contributed papers submitted to the Community Engagement Frontier: Diversity, Equity, and Inclusion Topical Group (CEF3). We have organized these topics based on the recommendations provided by Nord \textit{et al.} in the ``Culture change is necessary, and it requires strategic planning" letter of intent submitted as part of the Snowmass Community Planning Exercises \cite{LOI_StrategicPlanning}. The order of these recommendations does \textit{not} suggest prioritization. 
\vspace{-0.2in}

\subsection{CEF03 Recommendation 1.0 --- HEPA communities must employ the use of robust strategic planning procedures, including a full re-envisioning of science workplace norms and culture.}

\begin{recs}
    \rec{1.01}{Funding agencies and HEPA communities should prioritize community-related issues at the funding level.}{This might include the inclusion of community-related topics into safety parts of collaboration ``Operational Readiness Reviews," ``Conceptual Design Reviews," or similar documentation submitted to funding agencies. Funding agencies should provide clear and enforceable requirements for the advancement of DEI issues in grants, programs, and evaluations. Details in \cite{ClimateOfTheField}.}
    
    \rec{1.02}{Funding agencies should provide formal recommendations for institutions and collaborations for handling violations of their codes of conduct.}{This should include advice on handling community threats, removal of collaboration affiliates, leadership rights and responsibilities, and protections against legal liability for leadership that is responsible for that enforcement. This should also include advice on reporting to the funding agency itself; if there is no mechanism for reporting misconduct to a funding agency, that mechanism should be developed. Details in \cite{ClimateOfTheField}.}
    
    \rec{1.03}{Institutions and HEPA communities should develop effective reporting mechanisms and sanctions for egregious behavior.}{These institutions and communities should transparently describe those mechanisms in full for the benefit of all affiliates. Communities must be prepared to exercise those mechanisms. Future HEPA community codes of conduct should align with, and current codes of conduct should be reviewed upon new recommendations from funding agencies regarding enforcement and disciplinary measures. Details in \cite{ClimateOfTheField}.}
    \clearpage
    \rec{1.04}{Funding agencies should establish a dedicated Office of Diversity, Equity, and Inclusion to work with Program Officers.}{Funding agencies should establish the DEI Office to work with program officers to strategize and prioritize funding decisions and to develop equitable practices for the review processes. Details in \cite{MarginalizedCommunities}}
    
    \rec{1.05}{Funding and structural aid should be made available to develop ``Collaboration services'' offices at host laboratories.}{The community should prioritize the implementation of best practices networks across institutions and communities of physics practice. ``Collaboration services'' offices should provide HEPA collaborations and other physics communities of practice with the following: a) advice on legal and policy topics including those listed in F1.2, b) training in project management and ombudsperson training, c) logistical tools including facilitation of victim-centered investigation and mediation, d) resources and funding for local meeting accommodations, and other topics as described here in Section \ref{sec:recommendations}.  Details in \cite{ClimateOfTheField,AccessibilityInHEP, InformalSocializationInPhysicsTraining, PowerDynamicsInPhysics, ExcellenceAndEquityInPhysics}}
    
    \rec{1.06}{Institutions should have accessible, clear, robust, and flexible community-focused policies. Funding agencies should use their leverage to promote community-focused policies at funded institutions.}{Institutions should have accessible, clear, robust, and flexible community-focused policies for parental / family leave (for all genders) and vacation time. Funding agencies should require institutions that receive funding to implement policies on vacation time, parental \& family leave (for all genders), and health leave for all levels. Funding agencies should require institutions to prohibit confidentiality in settlements for egregious behavior (e.g. harassment); this promotes accountability and prevents known perpetrators from continuing to harm their communities. Details in \cite{LifestyleAndPersonalWellness, ClimateOfTheField}}
    
    \rec{1.07}{HEPA communities should support and take advantage of existing support structures and informational networks.}{Tools exist to support efforts to improve diversity and inclusion, as well as to address injustices in our communities. These include the American Association for the Advancement of Science (AAAS) Diversity and the Law program\cite{AAASDiversityAndTheLaw} which hosts resoures to enable promotion of legal and policy goals related to DEI. Knowledge like that collected by the American Institute of Physics’ National Task Force to Elevate African American Representation in Undergraduate Physics \& Astronomy (TEAM-UP) Project\cite{AIPTeamUpWebsite,TEAMUPreport2020} and the AAAS’s STEMM Equity Achievement (SEA) Change \cite{AAASSeaChange} should also be promoted. Details in \cite{PowerDynamicsInPhysics}.}
    
    \rec{1.08}{All community affiliates should reject harmful rhetoric and behavior related to work-life balance.}{This includes ``ideas around ‘lone geniuses’, the need for unhealthy work schedules, and the idea that sacrifice of personal wellness demonstrates your commitment to science"\cite{LifestyleAndPersonalWellness}. Senior scientists are responsible to ensure that they are managing their time and the time of those in their group properly to respect work-life balance (including meetings outside of working hours, or rotating meetings to accommodate varying time zones). Details in \cite{LifestyleAndPersonalWellness}.}
    \clearpage
    \rec{1.09}{Departments and institutions should have clear definitions of job responsibilities and ensure that they are funding all functions of the job.}{This includes any DEI work. Assessments should weight work in these areas equally and individuals should be awarded and/or recognized when they excel. Evaluation for employment should be based on carefully developed, public rubrics that include DEI, outreach, and service. Such rubrics should be created with considerable care and research-driven (e.g. if any of the criteria are biased in a way that would limit access or promotion of people who identify with an underrepresented group). Details in \cite{LifestyleAndPersonalWellness, ClimateOfTheField, InformalSocializationInPhysicsTraining}.}
    
    \rec{1.10}{Departments and institutions should reject the use of standardized exams in favor of holistic rubrics for admission.}{Evaluation for admission should reject the use of standardized exams and instead should be based on carefully developed, public rubrics, that are tailored to the department. Such rubrics should be created with considerable care and be research-driven (e.g. if any of the criteria are biased in a way that would limit access or promotion of people who identify with an underrepresented group). \cite{LifestyleAndPersonalWellness, ClimateOfTheField, InformalSocializationInPhysicsTraining}}
    
\end{recs}

\subsection{CEF03 Recommendation 2.0 --- HEPA communities must implement new modes of community organizing and decision-making that promote agency and leadership from all stakeholders within the scientific community.}
\begin{recs}
    \rec{2.01}{Funding agencies should facilitate Climate Community Studies.}{Studies should not be the responsibilities of individual communities. These studies should be informed by expertise in social and organizational dynamics. Details in \cite{ClimateOfTheField}.}

    \rec{2.02}{Reviews of community climate should include an evaluation of how leadership is selected within HEPA collaborations, as well as the valuation of sub-community contributions.}{This should include a expert-advised review of the assignment of high-impact analyses \& theses topics, convenership of working groups, and public-facing roles representing the collaboration such as spokespersons or analysis announcement seminars. Power dynamics within communities should also be evaluated, and should consider the impact that senior scientists can have especially on junior scientists of color. It should also include reviews of the participation of ``non-scientists" in community engagement and authorship, community perceptions of operations and service work, the development of onboarding and early-career networks, and implementation of policies toward equity in information sharing and software. Details in~\cite{ClimateOfTheField, PowerDynamicsInPhysics, InformalSocializationInPhysicsTraining, ExcellenceAndEquityInPhysics}.}
    \clearpage
    \rec{2.03}{Grant calls and assessments should include clear definitions of the tasks expected of PIs, including DEI related tasks, and provide grant funding for each.}{Alternatively, agencies could provide specific grants and awards for EDI and mentorship work. Agencies should ensure that they pay those on their grant review panels for their time. Details in \cite{LifestyleAndPersonalWellness}.}
    
    \rec{2.04}{Funding agencies should collect, analyze, and publish demographic information on grant proposals and funded grants.}{PI and funded \emph{and unfunded} researcher demographic information on grant proposals should be collected and used to track the effectiveness of these measures and are necessary to inform any additional policy changes needed to advance DEI policies and structures. Details in \cite{LifestyleAndPersonalWellness, InformalSocializationInPhysicsTraining}. }
    
    \rec{2.05}{Pay for student and postdoctoral researcher positions must increase.}{Graduate students and postdoctoral researchers should be paid at the level of their respective skill levels. Pay should include cost-of-living adjustments, relocation services, health coverage (including families), retirement savings, subsidized family housing. Trainees should not be taxed on fellowship money they do not receive as pay. These benefits must apply while students and their families are abroad on behalf of HEPA activities. Details in \cite{LifestyleAndPersonalWellness}.}
    
    \rec{2.06}{Collaborations should train members in standards in the field and offer mentorship programs to ensure that postdocs and students (especially from underrepresented groups) have additional support and resources.}{Mentorship programs should be research-driven and should make access to information as ubiquitous as possible. Mentors should help novices navigate the complicated landscape of the community, and care should be taken to address the ``untold rules", like non-academic career trajectories. Information sharing, especially about collaboration policies, procedures, and code-bases, should be evaluated from an equity lens. Details in \cite{LifestyleAndPersonalWellness, InformalSocializationInPhysicsTraining, ClimateOfTheField}.}
    \clearpage
    \rec{2.07}{The U.S. HEP community should maintain the current engagements and increase investments in Africa and Latin America to improve the reach of HEP in these regions.}{Funding agencies and international collaborations should acknowledge the disparity in economic capabilities of countries in Africa and Latin America compared to what is available in the U.S.. Funding agencies should support the development of HEP in these countries, should support and lead initiatives for more equitable contributions (e.g. membership and operations fees for participation in large collaboration, conference fee waivers and travel support to U.S. based meetings, etc). U.S. universities and research labs should encourage and support the participation of their personnel, faculties and research staffs in HEP education and research efforts of African and Latin American countries. U.S. institutes need to partner with Latin America and Africa in establishing bridge programs and supporting community members from Africa and Latin American to come to U.S. laboratories and universities for research experience programs. Collaborations and conferences should seriously consider decreasing or waiving membership and operations fees for participation and should provide financial assistance for travel to the U.S. Details in \cite{HEPInAfricaAndLatinAmerica}.}
    
    \rec{2.08}{Conferences should offer financial assistance to individuals with hardships.}{Conferences should offer limited travel grants through an application procedure overseen by an ethics group associated with the conference. To promote the engagement of under-resourced and early-career scientists, conferences should also strongly consider developing an application for sliding-scale / waiver for conference registration fees. Conferences should accommodate caregiving responsibilities by providing childcare onsite, or by supporting the travel of an accompanying person. In both situations, extra funding should be budgeted by the conference to fully or partially cover those costs. Details in \cite{AccessibilityInHEP}.}
    
    \rec{2.09}{All HEP activities must ensure that people with accessibility barriers are truly accommodated, with guaranteed, low-friction, dignified access to all aspects of the experience.}{All conferences, collaborations, universities, and labs should be made accessible to people with disabilities.  For example, conferences (including virtual meetings) should be announced with enough time to arrange accommodations for any individual needs, and organizers should plan to secure funding and book services far enough ahead of time. Accommodations should include both steno-captioning and ASL interpretation, which should be fully funded as part of the conference budget. Conferences should also be accessible to the blind / low-vision community, which may include screen-reader-accessible tools and ``color-blind-friendly" plots. Other accommodations include: seating or accessible access to amenities like check-in and meals; locating the conference in ADA-compliant buildings with no obstructions to seating, entrances / exits, or accessible pathways; quiet spaces; and designated contacts for troubleshooting accessibility.   An extensive list of recommendations can be found in \cite{AccessibilityInHEP}.}
\end{recs}

\subsection{CEF03 Recommendation 3.0 --- HEPA communities must engage in partnership with scholars, professionals, and other experts in several disciplines, including but not limited to anti-racism, critical race theory, and social science.}

\begin{recs}
    \rec{3.01}{Funding should be made available to both engage with and compensate experts in DEI, anti-racism, critical race theory, and social science.}{This can take the form of independent grants, but more effective would be the inclusion of climate-related topics into safety components of collaboration ``Operational Readiness Reviews,'' ``Conceptual Design Reviews,'' or similar documentation submitted to funding agencies. Details in \cite{ClimateOfTheField}.}
    
    \rec{3.02}{Experts should be adequately integrated into HEPA communities.}{This is motivated by the need to apply their expertise effectively, and should include collaboration communities. This may take the form of an official collaboration role like a non-voting member of a collaboration council. Details in \cite{ClimateOfTheField}.}
    
    \rec{3.03}{Community studies should be run by and receive advice from experts in sociology and organizational psychology.}{The tools used to evaluate the climate of HEPA need to be adequate, effective, and informative. These studies and accompanying expertise should be funded at the federal and institutional levels. They should include evaluation of leadership selection, development of junior scientists and their trajectories, and the existence of detrimental power dynamics that specifically affect underrepresented groups. Undesirable systems should be addressed with direct intervention. Details in \cite{ClimateOfTheField, PowerDynamicsInPhysics, InformalSocializationInPhysicsTraining}.}
    
    \rec{3.04}{Grant calls and assessments should include the advice of professionals in DEI and education.}{Such experts should review the entire process, including portfolios in their entirety, but with specific attention to mentorship and DEI plans. Experts should be paid for their time. Details in \cite{LifestyleAndPersonalWellness}.}
    
    \rec{3.05}{Identification of leaders within HEPA communities should be research-driven.}{HEPA organizations and institutions require leaders who will promote policies and practices that support underrepresented and historically marginized groups instead of favoring ``politics and convenience". Best practices have been developed by industrial \& organizational psychologists and are under studies at NSF (e.g.\cite{nsfADVANCEOrganizationalChange}). Details on necessary search practices can be found in \cite{ExcellenceAndEquityInPhysics}. }
\end{recs}

\subsection{CEF03 Recommendation 4.0 --- HEPA communities must develop effective programs to diversify the field and invest in creating structure.}

\begin{recs}
\rec{4.01}{Funding agencies and collaborations must allocate financial resources to create pathways and provide support for minoritized members in their academic and research pursuits.}{}

\rec{4.02}{Think strategically about how Community Engagement topics are integrated into other frontiers.}{This work is the work of \textit{all} HEPA community members, and should not be relegated entirely to an independent, volunteer-driven frontier.}
    
\rec{4.03}{Funding for Snowmass activities should include critical infrastructure for accessibility.}{This includes live captioning for all public events, and infrastructure for hybrid meetings to support those who cannot travel to attend workshops.}
\end{recs}





\bibliographystyle{JHEP}
\bibliography{Engagement/CommF03/cef3} 




\end{document}